\documentstyle[11pt]{article} 
\begin{document} 

\centerline{\Large Boltzmann's {\it H}-theorem and time irreversibility} 
\vskip 5pt 
\centerline{C. Y. Chen} 
\centerline{Department of Physics, Beijing 
University of Aeronautics} 
\centerline{ and Astronautics, Beijing, 100083, 
P.R.China} 
\centerline{Email: cychen@buaa.edu.cn} \vskip 5pt 
\begin{abstract} It is shown that the justification of the Boltzman 
$H$-theorem needs more than just the assumption of molecular chaos and the 
picture of time irreversibility related to it should be reinvestigated. 
\end{abstract} 
\vskip 5pt 

It is well-known that while the Newtonian formalism is time reversible the 
kinetic equations, based almost entirely on the Newtonian formalism, are time 
irreversible. This paradox has served as a serious topic for the century-long 
debate. Recent studies of non-equilibrium phenomena, such as those related to 
turbulence, transport and chaos, constantly remind us that a good 
understanding of time irreversibility is of crucial importance in terms of 
knowing the game nature plays. Keeping those things in mind, we shall here 
concern ourselves with explication and implication of time irreversibility in 
classical physics. It will be shown that the Boltzmann $H$-theorem and the 
picture of time irreversibility related to it involve mathematical problems. 

According to the standard theory\cite{reif}, the evolution of dilute gas 
consisting of hard sphere balls (referred to as particles hereafter) is 
described by the Boltzmann equation 
\begin{equation}\label{beq} \frac{\partial f_1}{\partial t}+ {\bf 
v}_1\cdot\frac{\partial f_1}{\partial {\bf r}}+ \frac{{\bf F}}m\cdot 
\frac{\partial f_1} {\partial {\bf v}_1}= \int \int d{\bf v}_2 d\Omega u 
\sigma (f_1^\prime f_2^\prime -f_1f_2). \end{equation} 
To show the time irreversibility of it, the theory introduces a 
function\cite{reichl} 
\begin{equation}\label{hfunction} H(t)=\int\int d{\bf r} d{\bf v}_1 f_1({\bf 
r},{\bf v}_1, t)\ln f_1({\bf r},{\bf v}_1, t), \end{equation} 
which can be recognized as a form of negative entropy. By substituting 
(\ref{beq}) into (\ref{hfunction}), the time derivative of $H(t)$ is, with 
external forces neglected, 
\begin{eqnarray}\label{tderivative} \frac{d H}{d t} =- \displaystyle{\int\int 
d{\bf r} d{\bf v}_1\left( \dot{\bf r} \cdot \frac{\partial f_1}{\partial {\bf 
r}} \right) [\ln f_1+1]}\qquad \qquad \qquad\quad \nonumber \\ +\int \int 
\int \int d{\bf r} d{\bf v}_1 d{\bf v}_2 d\Omega u \sigma (f_1^\prime 
f_2^\prime -f_1f_2) [\ln f_1+1]. \end{eqnarray} 
On the premise of that the distribution function vanishes for large $\bf r$ 
and ${\bf v}_1$, called the null boundary condition herein, we arrive at 
\begin{equation} \label{t1} \frac{d H}{d t} = \int \int \int \int d{\bf r} 
d{\bf v}_1 d{\bf v}_2 d\Omega u \sigma (f_1^\prime f_2^\prime -f_1f_2) [\ln 
f_1+1] . \end{equation} 
Since $f_2$, $f_1^\prime$ and $f_2^\prime$ describe the same gas, three other 
formulas similar to (\ref{t1}) can also be obtained; thus $dH/dt$ finally 
becomes 
\begin{equation} \label{t2} \frac{d H}{d t} =\frac 14 \int \int \int \int 
d{\bf r} d{\bf v}_1d{\bf v}_2 d\Omega u \sigma (f_1^\prime f_2^\prime 
-f_1f_2)\ln \frac {f_1f_2}{f_1^\prime f_2^\prime}\le 0, \end{equation} 
which is always less than zero except for gases that are in equilibrium. This 
conclusion, called the Boltzmann $H$-theorem, was, and still is, regarded as 
a great triumph of the Boltzmann theory, since it explained, rather 
generally, macroscopic time irreversibility in terms of microscopic laws. 

Notice that the derivation above specifically identifies particle-to-particle 
collisions as a mechanism responsible for time irreversibility. 
Interestingly, this identification, supposed to reveal the very secret of 
nature, confused, and continues to confuse, many scientists. The main reason 
lies in that the time irreversible $H$-theorem is, as has just shown, based 
on the properties of the Boltzmann collisional operator, while the Boltzmann 
collisional operator itself is based on the time reversibility of the 
Newtonian formalism. 

To make things more perplexing, an explicit theorem in textbooks, while based 
also on the Newtonian formalism and null boundary condition, tells us a 
different story\cite{harris}. The theorem, called the $\rho-S$ theorem in 
this paper, goes as follows. The entropy of a gas system is defined as 
\begin{equation} \label{entropy} S=-k\int d\Gamma \rho \ln \rho, 
\end{equation} 
where $\rho$ is the grand distribution of the system in the grand phase 
space, $\Gamma$-space. Differentiating $S$ with respect to time yields 
\begin{equation}\label{entropyt} \frac{d S}{dt} = -k \int d\Gamma (\ln 
\rho+1)\frac{\partial \rho}{\partial t}. \end{equation} 
Substituting Liouville's theorem $\partial \rho/\partial t + [\rho, H] =0$ 
into (\ref{entropyt}), we finally get, after a few mathematical steps, 
\begin{equation}\label{dsdt} \frac{dS}{dt}=0,\end{equation} 
which literally means that interactions between particles themselves, no 
matter what kinds of forms they take, are not responsible for time 
irreversibility. This theorem, though less popular 
than the Boltzmann one, is by no means surprising since it presents nothing 
but the time reversibility of Newtonian mechanics. 

The conflict between the two theorems above has been known for quite 
long\cite{some}. To 
resolve the difficulty, the conventional wisdom invoked the assumption of 
molecular chaos. It is widely believed that with help of molecular chaos 
the Boltzmann theory is justified at least in the practical sense\cite{cer}. 
However, being exposed to many interesting phenomena of gas dynamics, we are 
convinced that the Boltzmann theory needs more than just the assumption of 
molecular chaos. 

Let's first look at the time reversibility in mechanics.
Consider two identical particles (still distinguishable according to 
classical physics). The initial and final velocities of them are denoted by 
${\bf v}_1,{\bf v}_2$ and ${\bf v}_1^\prime,{\bf v}_2^\prime$ respectively. 
The usual concept of time reversibility states that if the collision ${\bf 
v}_1,{\bf v}_2 \rightarrow{\bf v}_1^\prime,{\bf v}_2^\prime$ is physically 
possible, then the inverse collision $-{\bf v}_1^\prime,-{\bf v}_2^\prime 
\rightarrow -{\bf v}_1,-{\bf v}_2$ is also physically possible, which is 
trivial and we discuss it no more. 

Then, we study the time reversibility concerning beam-to-beam collisions, 
from which the Boltzmann collisional operator is derived. To our 
great surprise, the time reversibility of this type does not exist at  
all: there is neither an intuitive one, nor a mathematical one. 

\hspace{-0.4cm} \setlength{\unitlength}{0.023in} 
\begin{picture}(100,76) \multiput(24,23)(-1,2){2}{\vector(2,1){22}} 
\multiput(24,52)(-1,-2){2}{\vector(2,-1){22}} \put(52,41){\vector(1,1){18}} 
\put(53,40){\vector(3,2){20}} \put(51,42){\vector(2,3){14}} 
\put(52,34){\vector(1,-1){18}} \put(53,35){\vector(3,-2){20}} 
\put(51,33){\vector(2,-3){14}} \put(50,2){\makebox(0,8)[c]{\bf (a)}} 
\multiput(145,39.6)(1,2){2}{\vector(-2,1){22}} 
\multiput(145,36.4)(1,-2){2}{\vector(-2,-1){22}} 
\put(171,59.5){\vector(-1,-1){18}} \put(174,53.5){\vector(-3,-2){20}} 
\put(166,63.5){\vector(-2,-3){14}} \put(171,16){\vector(-1,1){18}} 
\put(174,22){\vector(-3,2){20}} \put(166,12){\vector(-2,3){14}} 
\put(150,5){\makebox(0,8)[c]{\bf (b)}} \end{picture} 
\vskip-0.3cm 
\begin{center} 
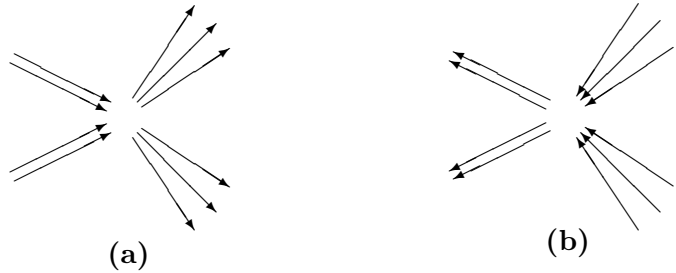
\begin{minipage}{10cm} {Figure~1: A candidate for time reversibility of 
beam-to-beam collision: (a) the original collisions; and (b) inverse 
collisions imagined. } \end{minipage} 
\end{center} 

Intuitively, we may consider two pictures sown in Fig.~1.
Fig.~1a shows that two particle beams at two definite velocities collide and 
the particles produced by the collisions diverge in the position and velocity 
space. Fig.~1b illustrates the opposite process, in which 
different converging beams collide and 
the produced particles form two beams, each of which has one definite 
velocity. In no need of discussion, we all find that the first picture makes 
sense in statistical mechanics, while the second one does not. 

In standard textbooks, the following mathematical definition of time 
reversibility has been employed\cite{reif}: 
\begin{equation}\label{equality} \sigma({\bf v}_1,{\bf v}_2 \rightarrow {\bf 
v}_1^\prime, {\bf v}_2^\prime) = \sigma ({\bf v}_1^\prime, {\bf v}_2^\prime 
\rightarrow {\bf v}_1, {\bf v}_2) ,\end{equation} 
where the cross section $\sigma({\bf v}_1,{\bf v}_2 \rightarrow {\bf 
v}_1^\prime, {\bf v}_2^\prime )$ is defined in such a way that, after 
collisions between a beam of type-1 particles at ${\bf v}_1$ and a type-2 
particles at ${\bf v}_2$, 
\begin{equation} \label{sigma1} N= \sigma({\bf v}_1,{\bf v}_2 \rightarrow 
{\bf v}_1^\prime, {\bf v}_2^\prime ) d{\bf v}_1^\prime d{\bf v}_1^\prime 
\end{equation} 
represents the number of type-1 particles emerging between ${\bf v}_1^\prime$ 
and ${\bf v}_1^\prime+d {\bf v}_1^\prime$ per unit incident flux and unit 
time, while the type-2 particle emerges between ${\bf v}_2^\prime$ and ${\bf 
v}_2^\prime+d {\bf v}_2^\prime$; and the cross section $\sigma({\bf 
v}_1^\prime ,{\bf v}_2^\prime \rightarrow {\bf v}_1, {\bf v}_2)$ is defined 
in the same manner. 

\hspace{5.8cm} \setlength{\unitlength}{0.016in} 
\begin{picture}(200,142) \put(-120,85){\vector(3,-1){84.5}} 
\put(-120,85){\vector(1,1){27.5}} \put(-120,85){\vector(1,0){56.5}} 
\multiput(-91.5,113.5)(1.5,-1.5){36}{\circle*{1.2}} 
\put(-86.5,113.5){\vector(1,-1){56.5}} \put(-86.5,113.5){\circle*{3}} 
\put(-106,109){\makebox(35,8)[l]{${\bf v}_1$}} 
\put(-58,53){\makebox(35,8)[l]{${\bf v}_2$}} 
\put(-90,77){\makebox(35,8)[c]{${\bf c}$}} 
\put(-68,80){\makebox(35,8)[c]{${\bf u}$}} \put(-80,30){\makebox(0,8)[c]{\bf 
(a)}} \hspace{-2cm} \put(98,30){\makebox(0,8)[c]{\bf (b)}} 
\put(45.5,85){\vector(1,0){56.5}} 
\multiput(103,85)(-1.5,-1.5){19}{\circle*{1.2}} 
\multiput(103,85)(1.5,1.5){19}{\circle*{1.2}} 
\put(45.5,85){\vector(3,1){84.5}} \put(45.5,85){\line(5,1){92}} 
\put(45.5,85){\vector(2,1){72.5}} \put(136.5,103.4){\vector(4,1){1}} 
\put(45.5,85){\vector(4,3){52.5}} \put(45.5,85){\vector(1,-1){28.5}} 
\put(45.5,85){\vector(1,1){28.5}} \put(56,111){\makebox(35,8)[l]{$({\bf 
v}_1^\prime)$}} \put(88,127){\makebox(35,8)[l]{$({\bf v}_1^\prime)$}} 
\put(113,124){\makebox(35,8)[l]{$({\bf v}_1^\prime)$}} 
\put(132,114){\makebox(35,8)[l]{${\bf v}_1^\prime$}} 
\put(139,102){\makebox(35,8)[l]{$({\bf v}_1^\prime)$}} 
\put(61,51){\makebox(35,8)[l]{${\bf v}_2^\prime$}} 
\put(69,77){\makebox(35,8)[c]{${\bf c}$}} \put(76,65){\makebox(35,8)[c]{$u$}} 
\put(133,50){\makebox(35,8)[l]{$S$}} 

\put(143.00, 85.00){\circle *{1.2}} \put(142.78, 89.18){\circle *{1.2}} 
\put(142.13, 93.32){\circle *{1.2}} \put(141.04, 97.36){\circle *{1.2}} 
\put(139.54, 101.27){\circle *{1.2}} \put(137.64, 105.00){\circle *{1.2}} 
\put(135.36, 108.51){\circle *{1.2}} \put(132.73, 111.77){\circle *{1.2}} 
\put(129.77, 114.73){\circle *{1.2}} \put(126.51, 117.36){\circle *{1.2}} 
\put(123.00, 119.64){\circle *{1.2}} \put(119.27, 121.54){\circle *{1.2}} 
\put(115.36, 123.04){\circle *{1.2}} \put(111.32, 124.13){\circle *{1.2}} 
\put(107.18, 124.78){\circle *{1.2}} \put(103.00, 125.00){\circle *{1.2}} 
\put(98.82, 124.78){\circle *{1.2}} \put(94.68, 124.13){\circle *{1.2}} 
\put(90.64, 123.04){\circle *{1.2}} \put(86.73, 121.54){\circle *{1.2}} 
\put(83.00, 119.64){\circle *{1.2}} \put(79.49, 117.36){\circle *{1.2}} 
\put(76.23, 114.73){\circle *{1.2}} \put(73.27, 111.77){\circle *{1.2}} 
\put(70.64, 108.51){\circle *{1.2}} \put(68.36, 105.00){\circle *{1.2}} 
\put(66.46, 101.27){\circle *{1.2}} \put(64.96, 97.36){\circle *{1.2}} 
\put(63.87, 93.32){\circle *{1.2}} \put(63.22, 89.18){\circle *{1.2}} 
\put(63.00, 85.00){\circle *{1.2}} \put(63.22, 80.82){\circle *{1.2}} 
\put(63.87, 76.68){\circle *{1.2}} \put(64.96, 72.64){\circle *{1.2}} 
\put(66.46, 68.73){\circle *{1.2}} \put(68.36, 65.00){\circle *{1.2}} 
\put(70.64, 61.49){\circle *{1.2}} \put(73.27, 58.23){\circle *{1.2}} 
\put(76.23, 55.27){\circle *{1.2}} \put(79.49, 52.64){\circle *{1.2}} 
\put(83.00, 50.36){\circle *{1.2}} \put(86.73, 48.46){\circle *{1.2}} 
\put(90.64, 46.96){\circle *{1.2}} \put(94.68, 45.87){\circle *{1.2}} 
\put(98.82, 45.22){\circle *{1.2}} \put(103.00, 45.00){\circle *{1.2}} 
\put(107.18, 45.22){\circle *{1.2}} \put(111.32, 45.87){\circle *{1.2}} 
\put(115.36, 46.96){\circle *{1.2}} \put(119.27, 48.46){\circle *{1.2}} 
\put(123.00, 50.36){\circle *{1.2}} \put(126.51, 52.64){\circle *{1.2}} 
\put(129.77, 55.27){\circle *{1.2}} \put(132.73, 58.23){\circle *{1.2}} 
\put(135.36, 61.49){\circle *{1.2}} \put(137.64, 65.00){\circle *{1.2}} 
\put(139.54, 68.73){\circle *{1.2}} \put(141.04, 72.64){\circle *{1.2}} 
\put(142.13, 76.68){\circle *{1.2}} \put(142.78, 80.82){\circle *{1.2}} 
\end{picture} 
\begin{center} 
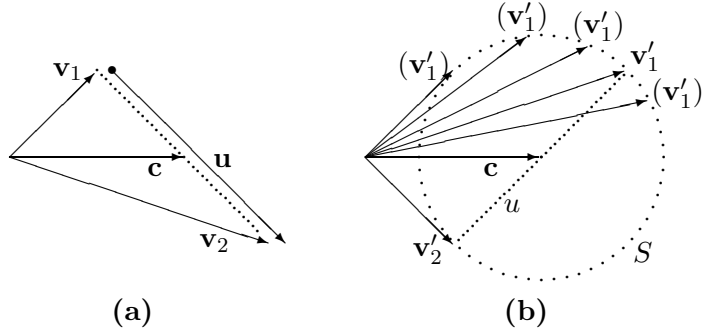
\begin{minipage}{10cm} {\vskip -1.9cm Figure~2: Constraints on the final 
velocities of scattered particles. (a) ${\bf v}_1$ and ${\bf v}_2$ 
predetermine ${\bf c}$ and $u=|{\bf u}|$. (b) ${\bf v}_1^\prime$ and 
${\bf v}_2^\prime$ have to fall on the shell $S$ of diameter $u$. } 
\end{minipage} 
\end{center} 
\vskip -0.4cm 

An unfortunate fact with the time reversibility (\ref{equality}) is that
the cross section in it is mathematically ill-defined. For the collisions 
between two beams with ${\bf v}_1$ and ${\bf v}_2$ respectively, the 
energy and momentum conservation laws imply that ${\bf v}_1^\prime$ and 
${\bf v}_2^\prime$ satisfy 
\begin{equation} \label{csv} {\bf v}_1^\prime+ {\bf v}_2^\prime ={\bf 
v}_1+{\bf v}_2 \equiv 2{\bf c} \quad {\rm and} \quad |{\bf v}_2^\prime- {\bf 
v}_1^\prime| = |{\bf v}_2-{\bf v}_1| \equiv u, \end{equation} 
where ${\bf c}$ is the center-of-mass velocity and $u$ is the relative speed. 
Fig.~2a shows that ${\bf c}$ and $u$ are determined by ${\bf v}_1$ and 
${\bf v}_2$, while Fig.~2b shows that $\bf c$ and $u$ impose constraints on 
${\bf v}_1^\prime$ and ${\bf v}_2^\prime$. Referring to the figures, we 
find that 
${\bf v}_1^\prime$ and ${\bf v}_2^\prime$ must fall on a spherical shell 
$S$ of diameter $u=|{\bf u}|$ in the velocity space, called the 
energy-momenta shell herein. By adopting this notion, two misconcepts 
associated with (\ref{sigma1}) can be unveiled immediately. The first is that 
after $d{\bf v}_1^\prime$ is specified, specifying $d{\bf v}_2^\prime$ in 
(\ref{sigma1}) is a work overdone. The second is that the cross section 
should be defined in reference to an area element on the energy-momenta shell 
rather than in reference to an arbitrary velocity volume element.
If we insist on doing the latter,
the resultant `cross section' can equal any value from zero 
to infinity, depending on how $d{\bf v}_1^\prime$ 
encloses the shell and how $d{\bf v}_1^\prime$ approaches zero. To 
see it, imagine that $d{\bf v}_1^\prime$ is a cylinder centered 
on and perpendicular to the shell. When $d{\bf v}_1^\prime$ 
becomes slimmer and slimmer, $\sigma\rightarrow 0$; when 
$d{\bf v}_1^\prime$ becomes shorter and shorter, 
$\sigma\rightarrow \infty$\cite{chen1}. 

The above discussion has shown that, the immediate problem  
is not that we cannot build up a theorem that produces the time 
irreversibility assumed by the existing formalism but that we cannot build up 
a theorem that produces the time reversibility assumed by the existing 
formalism; and the problem is that of mathematics and has nothing to do with 
how do we employ subtle physical assumptions. 

It is now in order to return to our original subject and comment on physical 
mechanisms responsible for time irreversibility. According to information 
theory, the increase of entropy represents the destruction of information. 
For a classical gas, the information contained is nothing but the aggregate 
of all initial conditions. Whenever and wherever some of the initial 
conditions are erased, mechanisms of time irreversibility must be at work. 
Armed with this concept, we 
realize that chaos theory, as well as some analyses concerning averaging 
molecular motions, indeed promises to account for time irreversibility to 
some degree. Nevertheless, for purposes of this paper, boundary effects will 
be the subject of our discussion. 

In view of that the $\rho-S$ theorem is time reversible and the null boundary 
condition related to it is just an assumption, it is arguable that time 
irreversibility may arise from realistic interaction between boundaries and 
particles. 

Examine the situation in Fig.~3a, where a cuboid box with perfectly flat 
walls is stationary and all particles in it move with a definite speed $v$ 
rightward or leftward. To make the examination simpler, let all particles be 
rather small (or, the gas be rather dilute) so that particle-to-particle 
collisions are presumably negligible. Under these we find, 
if particle-to-boundary collisions are assumed to be perfectly elastic, 
the system will remain in the original state for quite long; whereas, if 
realistic boundary conditions are allowed to apply, the system will approach 
its equilibrium rather realistically. 
In Fig.~3b, a particle colliding with the 
boundary loses its memory of initial condition, at least partly, and spreads 
with various velocities according to a statistical law, in which fluctuation 
and dissipation must get involved\cite{kogan}. By setting the temperature $T$ 
of the walls uniform and such that $3\kappa T/2=mv^2/2$, it is seen that the 
gas entropy increases with no macroscopic energy exchange between the gas and 
boundary. 

\hspace{-0.1cm} \setlength{\unitlength}{0.023in} 
\begin{picture}(100,70) 

\put(30,30){\framebox(40,30){}} 

\put(50,40){\circle*{1.5}} \put(50,40){\vector(-1,0){5}} 
\put(62,46){\circle*{1.5}} \put(62,46){\vector(-1,0){5}} 
\put(42,39){\circle*{1.5}} \put(42,39){\vector(-1,0){5}} 
\put(42,50){\circle*{1.5}} \put(42,50){\vector(1,0){5}} 
\put(35,33){\circle*{1.5}} \put(35,33){\vector(1,0){5}} 
\put(55,36){\circle*{1.5}} \put(55,36){\vector(1,0){5}} 
\put(52,47){\circle*{1.5}} \put(52,47){\vector(-1,0){5}} 
\put(58,55){\circle*{1.5}} \put(58,55){\vector(1,0){5}} 
\put(40,55){\circle*{1.5}} \put(40,55){\vector(-1,0){5}} 

\put(50,16){\makebox(0,8)[c]{\bf (a)}} \hspace{-15pt} 
\put(150,45){\circle{25}} \put(162.5,30){\line(0,1){30}} 
\multiput(162.5,31.5)(0,3){10}{\line(1,0){3}} 
\put(162,45){\vector(-1,0){24.5}} \put(162,45){\vector(-2,-1){19}} 
\put(162,45){\vector(-1,-2){4.5}} \put(162,45){\vector(-2,1){19}} 
\put(162,45){\vector(-1,2){4.5}} 

\put(127,37){\circle*{2}} \put(127,37){\vector(4,1){8}} 
\put(146,16){\makebox(0,8)[c]{\bf (b)}} \end{picture} 
\begin{center} 
\begin{minipage}{10cm} {\vspace{-1.2cm} Figure~3: 
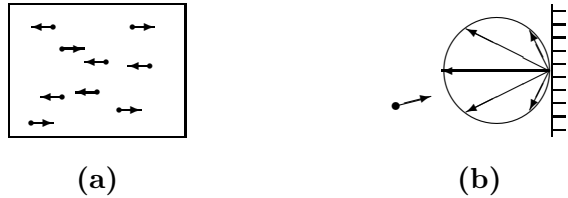
(a) Particles move rightward or leftward inside a box. (b) 
Schematic of how a particle collides with a boundary.} 
\end{minipage} \end{center} 

The relevance of the aforementioned effect can be verified quantitatively. 
For a dilute gas, we can simulate the relaxation process with help of 
certain empirical laws of particle-to-boundary collisions. It is very easy 
to see that while the $H$-theorem predicts much longer relaxation times 
for such gas 
($\tau\propto f^{-2}$) the undeterministic boundary specification may 
yield results consistent with those observed in nature. 

Our discussion also suggests that after colliding with a boundary a 
particle has to be regarded as part of particle beam in view of that the 
later motion of it can be known only in probability. As has been revealed, if 
such a beam meets with other beams, the consequent collisions should, in 
turn, be considered as being of time irreversibility. 

More investigations reveal more interesting things\cite{chen,chen2}, of which 
one is that realistic boundaries can not only erase information, but, in many 
cases, create information. It is no wonder that so many striking phenomena in 
fluid experiments occur around boundaries. 

Helpful discussions with professors Hanying Guo, Ke Wu and Qiang Chen are 
greatly appreciated. This paper is supported by School of Science, BUAA, PRC.

\end{document}